# 8
# *Educational research and teaching strategies in the digital society: a critical view*

*José Gómez Galán*
*Universidad de Extremadura (Spain)*
*Universidad Metropolitana. Sistema Universitario Ana G. Méndez (Puerto Rico)*
*jgomez@unex.es - jogomez@suagm.edu*

## 1. INTRODUCTION

Educational research throughout the twentieth century was too biased in different specialties that have had extremely defined methodologies and favorite models. From classic educational psychology, biology education, philosophy of education, sociology of education, economics of education, history of education, educational theory, etc., to specialties within the scope of teaching approach or new frameworks of educational technology. There have been multiple compartments in which science education has been fragmented with their corresponding research models (Gómez Galán, 2015a).

This situation has reached today. Thus, this is the first problem we have to face: educational research hasn't been tackled as a whole, from a global perspective, in the pursuit of dialogue among the multiple disciplines that were taking shape in the context of an excessive specialization which, academically, opened the way into the twentieth century, and in so many cases made it impossible to approach educational problems in many different areas.

Educational research as part of research in the social sciences and, overall, science and technology, has always been the subject of extensive discussions (Phillips & Burbules, 2000; Biesta & Burbules; 2003; Anyon, 2008; Johnson & Christensen, 2014; Pring, 2014; Wiersma & Jurs, 2014; Ponce & Pagán, 2015; Mertler, 2015; Gómez Galán & Sirignano, 2016; Chizzotti, 2016; Ponce, Pagán & Gómez Galán, 2016; Vigano, 2016; Ponce, Gómez Galán & Pagán, 2017; Beneito-Montagut, 2017). Wallerstein (2001) held that all concepts and analytical framework derived from research in the social sciences, among which education is, needed a rigorous critical examination so they could really fulfill their potential to adequately describe what society needs. Of course we would add that it goes without saying that any research that lacks deep interpretations of the fact analyzed to provide a critical dialogue with all those focusing on the same object could not be defined as such. It would only set out data more or less useful, but it could not be called *research*.

In the world of education the number of variables is huge. Thus, multidisciplinary studies are needed, of various methodologies, to approach the knowledge of the problems (Gómez Galán, 2005). And not just enough field work, currently focused too often on case studies difficult to apply in different contexts, but the development of theoretical *constructs* necessary to enable us to not only know what we are doing today in the field of education, but to ask what we should do for progress and social development in the search for a better world, and facing problems holistically. However, it is necessary to reunify the sciences of education in all thematic and methodological dimensions. The new digital society in which we live is creating new problems in education that can only be tackled from a multidisciplinary perspective. The education world is being transformed by the impact of the many changes that are taking place in the social, economic, political and cultural contexts occurred by the rise of information and communication technology (ICT). We must redesign a new education for a new time, and research is essential for this purpose. Therefore we must consider educational research as a reflection of what we want education to become. We have always argued that civilization has been built through education, and only this is what allows us to progress and move forward. From this derives the importance of research in this field of knowledge.

But a probably greater problem is to be faced, hugely increased in recent decades. Precisely the possibilities of managing high volumes of data by ICT led educational researchers, having always been dazzled by the quality of scientific methodologies in the field of experimental sciences, to make their commitment for using complex methods and quantitative techniques, more focused in many cases on the fact that in the essence. The splendor of the statistical calculations generalized the presence of basically quantitative studies, in many cases more prevalent than is indicated, with negligible effect on the advance of what education really is, and irrespective of released anecdotal data, which are more suitable to satisfy a curiosity and / or simply to the attainment of academic achievement. Sometimes research models that are set out and applied in multiple case studies are clearly unnecessary as plenty of samples to reach conclusions are already available. But the theoretical magnificence of scientific rigor – it must be stressed: with professional connotations for researchers rather than real use in its

application – often deserves credit far above what that research is actually providing us with. Even studies that would be fully justified by qualitative methods are transformed into quantitative or, to say the least, mixed, with the deductive process in the service of the method rather than the objectives (Gómez Galán, 2015a).

## 2. CURRENT CONTEXT OF EDUCATIONAL RESEARCH

Quantitative methodologies, qualitative and mixed methodologies, positivist and naturalists paradigms, experimental and quasi-experimental designs, action-research, statistics, IT developments, etc., are, as Bisquerra (2009) stated, common words in the current language of educational research. Sometimes we are not carried away by fashions, methods that are successful in a particular field and are generalized to a whole category without being accompanied by a sober reflection and on no justifiable basis. And conversely, techniques that were traditionally successful and offered excellent results are vilified, cornered and taken into oblivion (Gómez Galán, 2016).

But there is an even greater problem in all these dynamics, already referred to and in which we must stop as it is scarcely mentioned: the fact that the result of these investigations - regardless of its scope - is restricted to a merely academic and professional realm, without any communication of results - which would involve putting them into practice - to the educational community, teaching professionals, families, students, who are the actual protagonists of what we understand as education. Walker (1985) warned about the serious problems of communication that existed in educational research, and although paradoxically we are in a hyper-connected world, far from being solved they have increased almost to infinity.

In our view there is, no doubt, an explanation for it: the condition of university professional of educational research. Far from having absolute freedom to focus their efforts on the real problems that affect education today – which, as mentioned above are the main problems that surround our society - educational researchers are forced to investigate in order to publish. University professionals are the only ones who are trained to meet educational methodologies and, consequently, are involved in the field of education. But the accountability is towards their institutions, not society, and they do have to be assessed and measured in terms of the theoretical quality of their publications, which are conditioned by impact factors and *rankings* of powerful publishing and media trusts, with ramifications in business, economic areas and, no doubt, political ones.

The time when the education professional will also be a researcher is still very far. He who researches in education is a professional, above all, of the research itself. Certainly in Higher Education both dimensions merge, but not on other educational levels, the most important ones for the future of society. Research in Early Childhood Education, Primary or Secondary Education are not performed by educators who daily work and live by them, but for those professionals of university research whose methodologies are conditioned by a final product, the *article*, which will be subordinated on them by the impact indexed journals.

And no matter the why or what or the target. What matters is how, the research method. For an article on education today will be asked primarily to resemble an article of experimental science, that is, worked out by professional researchers, and for this purpose a methodology as similar as possible will be required. The problem, question or concern that affects our society - what will really transform and improve our world - will remain in the background. Let's just look globally at the issues of educational articles published in scientific impact journals, and check whether they really provide what today's education requires. Or, conversely, they are conditioned by bibliometric parameters which oppress and condition the freedom of the researcher, since the article assessors, let there be no mistake, take into account mainly the method. Methodologically perfect articles, but in many cases devoid of meaning and practical implementation. In this context the true professionals of education on each educational level have little chance to investigate. And if they do they will be conditioned by the investigative methods. If they do not use the suitable ones - and in many cases we refer to those methods currently in-style- they won't obviously ever see their contributions published, however relevant they may be. They will encounter problems to raise and design their research projects (Gómez Galán, 2015a).

## 3. EDUCATIONAL RESEARCH DEDICATED TO EDUCATION

Thus, training in this area is absolutely necessary for any educational professional (Pring, 2004; Deem & Lucas, 2006; Glasserman & Ramírez, 2015; Gómez Galán & Lacerda, 2012; Cobos, Gómez Galán, & López Meneses, 2016). It is therefore advisable that those who carry out their daily work in that particular level should address these field studies. It should not be forgotten that those professionals own valuable information on most occasions inaccessible by other research methods: their daily experience. This experience must necessarily be systematized and organized if we really want to achieve the objectives. Stenhouse (1980) was aware of the importance of establishing procedures closer to educational logic- mainly focused on values- than those offered by models of standard objectives of institutions and bureaucracies focused on financial yield and procedures measurable by quantitative parameters. This author also underlined the importance of teachers, schools and educators in general since they are the mainstay of research and responsible for transforming the curriculum (Stenhouse, 1981). Unfortunately today, due to the above circumstances, the situation is even worse than this author already pointed out a few decades ago.

What solutions can we find in this regard? Necessary steps must be taken to transform this reality which in no way is helping to improve education and consequently society. Some of the essential ones will be pointed out: (a) enhancing teacher training, at all levels, in educational research from a scientific perspective obviously, and also making measures towards including these professionals within the multidisciplinary context where education is located.; (b) a major issue is to highlight the importance of methods, stressing their potential to solve educational problems that need to be solved. Also, to achieve goals that may improve education; (c) supporting primary and secondary school teachers who want to innovate and investigate. Help them put into practice the results of their studies and as well as their wide dissemination if they have been successful; and (d) ending the tyranny of bibliometric indicators to assess educational research, which conditions its outcome. Today it is really to know if research is carried out either seeking an educational and social benefit or to achieve different academic and professional goals. It is evident that on many occasions it is unfeasible for both purposes (Gómez Galán, 2015a).

Working with humans involves the study of so many variables that the methods typical of experimental sciences cannot be adopted without any further research. And these are the methods which are mainly present in journals indexed in the databases referred to by institutions and administration. In many cases it may be more useful for improving the teaching-learning, for example, a study based on the life of a teacher and what he did in his professional career -using traditional methods typical of the Humanities and not of the Social Sciences - . As stated above, this may be more helpful than a research, for instance, on the integration of ICT in a particular region using a sample of 4500 people through a questionnaire of 120 items analyzed by all kinds of parametric and nonparametric tests statistics. In the present context the first study would be difficult to be published and, if so, it would be offered in an environment of very low diffusion. And that teacher might carry out amazing innovations extremely useful for their colleagues. The second study, however, would not have many problems to be published in a journal of high impact but what he might offer would quite probably be something long confirmed and corroborated by dozens of similar studies.

But this is the context in which today, unfortunately, educational research is moving. Nevertheless, there is nothing more typical of research that thinking: a human characteristic and what should mark the basis of research. It should be kept in mind that we investigate to answer new questions. It is necessary to humanize educational research. Inspire to investigate all what is necessary for the purpose. We believe it is time to make educational research more humane, even though we use the most comprehensive statistical methodologies in the service of better education, or to solve serious problems we face or to give answers to what we really need and are compelled by. The target is not to be sought from the method, and in the service of it. Rigor is essential, no doubt, but for something that makes sense. We must not forget that the Sabbath was made for man, not man for the Sabbath. We live in an academic context, in the service of improving education and, consequently, society, in which some tremendously questionable standards – performed many times by people who have never set foot in a classroom – are becoming extremely relevant. We'll make ours the words of Bertrand Russell (1950) when he said that educators and teachers, above other professionals, are the custodians of civilization. To which we would add something else: they are the builders of civilization. Hence is derived the crucial importance of research in the field of education.

## 4. CONECTION BETWEEN EDUCATIONAL RESEARCH AND TEACHING STRATEGIES

Teaching strategies have always been understood as the set of educational decisions a teacher must make to facilitate the personal development of students and, from an educational perspective, this would have an impact especially on teaching-learning processes (McGonigal, 2005; Killen, 2007; Rivero, Gómez Zermeño & Abrego, 2013; Schmeck, 2013; Gómez Galán, 2015b). It is, therefore, an extremely far-reaching, delicate process. Adopting useful and versatile strategies in education contributes decisively to the quality of learning. Teaching strategies are the culmination of the educational process, which allows us to achieve the objectives and makes the student acquire the skills and abilities needed. All other key elements in education (educational policy, collaboration and involvement of families, teacher training, etc.) could fail if the work in the classroom is not suitable. Even the most innovative teaching methods, and think for example in the current processes of e-learning, in which the teacher becomes a coach rather than a transmitter of contents, didactic dynamics pursuing learning must always be adapted to the needs of student, whose involvement must be fully active; nevertheless, the design, creation and implementation of relevant teaching strategies must be the work of education professionals (regardless of the subsequent participation of students in teaching-learning, which is assuming much more importance nowadays than in traditional models with the new technological means) (Gómez Galán, 2015b).

From the teaching perspective the teacher should be the main driver, counselor, manager and developer of the leading educational dynamics in educational processes. We are dealing with teaching professionals. It is true that, throughout history, high-quality, efficient self-learning processes have been produced. In some circumstances this is not only desirable but also essential. But when we are talking about formal education, integrated in a state-run education system, in a context of highly complex massification conditioned by structural and legislative frameworks, the teacher's teaching ability is certainly decisive. In this educational level the employment and development of appropriate teaching strategies, especially if they are innovative and original, and ultimately attractive for the student, are crucial to the success and quality of education.

The classroom atmosphere is substantially affected by the use of teaching strategies. Undoubtedly, the students will be the final recipients and beneficiaries (Kintsch & Van Dijk, 1978). When someone has confidence in their teacher, and collaborates enthusiastically in the educational proposals they organize, it is clear that the classroom atmosphere will be excellent. It is true that in regular school dynamics certain automatic strategies can be performed (it has already been referred to) without the presence of control or preplanning. However, this can lead to a certain monotony that might turn to be counterproductive. The teacher must continually innovate and seek the attention of students. Therefore, an intensive tracking of procedures providing any unexpected or inconvenient situation, and always at the service of the interaction, will always allow a didactic and productive use of the time available for each class. In this regard, teaching strategies that have been successful should be advised and, in the case of innovations, it is essential to take very good account of the results obtained just in case improvements or changes were required.

To classify the teaching strategies we can use the classical proposal by Joyce and Weil (1980), who group them into four models (although they could be extended, depending on the educational context), in relation to their effect on student behavior: (a) *Information processing*: those seeking conductive observable improvements in a long period of time; They are more effective in the specific operational stage. Well find: training basics, advance organizers, inductive-discovery strategies, etc.; (b) *Personal* (synectic, non-directive, etc.): Personal-oriented models include strategies towards positive self-concept and inter-group attitude improvement; (c) *Social*: those related to cooperative behavior, reduction of intergroup tensions, feelings of empathy and antisocial behavior improvement (we can mention democratic learning, role plays, etc.); and (d) *Behavioral* (reinforcement, self-control, etc.) directly related to learning skills and performance.

## 5. TEACHING STRATEGIES AND ICT: INTEGRATING RESEARCH IN THE 21ST-CENTURY CLASSROOM

The great irruption of ICT (information and communications technology) in society does not alter the traditional classification. It should be kept in mind that, if used in the classroom as a resource or teaching assistant, they are instruments at the service of teachers and students, seeking an improvement in the process of teaching and learning, but always considering that the human dimension is essential. On the other hand, integrated as an object of study, they would be elements of our society subject to a critical study and analysis. Therefore we speak of content that can be treated perfectly in the field of classical teaching strategies. However, the current situation in which we are engaged in educational contexts makes it very complex to work with appropriate strategies. There are too many barriers that hinder it. Distance from theory to practice is increasingly growing. In this sense, for example, ICT are decisively important. Today children and youth belong to a digital world in which they constantly receive stimuli outside the scope of educational environments, which makes teacher performance highly difficult. Therefore teachers are forced to adapt to this reality but, at the same time, they have to distance themselves from instructional processes. We must not forget that the use of these tools on a social level is basically for entertainment, after having outdated initial management features. Therefore the professional attitude of teachers should be trying to optimize their benefits and minimize their drawbacks, getting them to be used in teaching strategies at the service of training and educational growth (Gómez Galán, 2015b & 2017).

There is no doubt that today's teachers are being demanded many more challenges than the can assume. And among these challenges they are required by governments to introduce ICT in educational processes (sometimes with a shoehorn). It is assumed that these media will be essential in tomorrow's world (they are already today) but precisely the mistake is not recognizing that the instrumental use by children and young people is at least equal, if not superior, to their teachers. Therefore, from this point of view its integration is unnecessary, quite the opposite as to what an exhaustive analysis would mean as an object of study, examining their presence and significance in our world, in order to create critical attitudes towards their power of influence.

But we insist that the forced integration of ICT in teaching strategies used in the classroom can be extremely dangerous. The actual status of teachers is that, on many occasions, they are immersed in a far from conductive context to achieve goals that require new directions in the dynamics of teaching and learning environment. Of course, in a stressful classroom atmosphere any innovative teaching proposal may fail if the possibilities and limitations are not previously assessed. The mere use of ICT does not guarantee optimal results, and teaching innovations may well come from other initiatives and areas. The experience of the teacher, his intellectual capacity and his mastery of the art of teaching may undoubtedly be much more effective than the mere application of theoretical proposals from contexts outside the daily educational reality, which incidentally is packed with contributions and studies attributing ICT almost magical educational virtues. It is becoming increasingly necessary to distinguish between teacher and educational researcher when, paradoxically, they would have to be synonymous.

## 6. GENUINE EDUCATION IN THE DIGITAL SOCIETY

Teaching strategies, therefore, not only have to be at the service of the contents but, in parallel, they have to address the challenge of achieving improved personal relationships as well as a favorable environment inside the classroom (Fraser & Walberg, 2005; Urdan & Schoenfelder, 2006; López Meneses & Gómez Galán, 2010; Cohn & Fraser, 2016). They must globally contribute not only to a process of instruction but, above all, the process of *education*. Undoubtedly, to tackle this goal successfully implies adequate *pedagogical* training of teachers. Of course we refer to a context not exclusively educational, but also based on the conviction that every educational act in the classroom should imply personal and human growth of students. To *educate* is the most wonderful of jobs, but it involves putting faith, above all, in work because, regardless of the immediate gratification obtained when it is loved, it is the only profession in which you are not working for the present (as would a lawyer, a doctor or a computer specialist) but for the future since only after one or two generations the teacher, and by extension society, reaps the rewards.

Unfortunately, it is uncommon to see a truly *pedagogical* training, and understand teaching strategies as part of human growth and not just as a set of aseptic professional techniques to be applied as in any other workplace. Moreover, training is not prevalent in our society nor does it characterize the syllabuses of future professionals of education. Particularly in certain academic standards it is not uncommon to find teachers who make the genuinely *educational* processes highly relative to the detriment of the acquisition of knowledge on different areas or subjects of their specialty. They seem to be (and without denying the importance of the examples below) more concerned that a student learns quadratic equations, or the economic policy of Philip II, or the characteristics of the poetic prose of Juan Ramon Jimenez, rather than developing a comprehensive training and the personal growth of each student according to their abilities. Clearly, the acquisition of this knowledge will contribute to it, without any doubt, but this is just as instruction, one element of an educational process. In this context, in which academic results prevail and in which teaching and learning (on many occasions clearly separated) are oriented towards a summative evaluation to measure - and sometimes not so precise - school performance, there is sometimes little scope for genuine *education*.

The main aim by no means should be that of making students memorize or assimilate the contents described, and learning teaching strategies employed must be in the service of this learning. What is really important and vital is that students, in the examples noted, marvel that the whole universe is governed by mysterious laws and that they can be glimpsed through mathematics. Also, the importance history has to improve the future or how we can learn from our mistakes and not repeat them; how to read Juan Ramon Jimenez, besides allowing us to enjoy a pure, extraordinary Spanish. All of this leads directly to the heart of human beings and everything good and beautiful that can arise from it.

## 7. CONCLUSIONS

We are talking about training to learn to love the teaching profession, helping others grow while we ourselves grow and help improve society in the search for a fairer world. Unfortunately, teacher training is increasingly limited to vocational training, of a purely technical nature, and in many cases demands a huge motivation from all those who believe that education should be something very *different*. *Educator* and *transmitter of knowledge* are antagonistic terms. *Educator* (the real teacher) is one who makes *love* knowledge, and in search of this beauty they should be trained. This training will not only provide teachers with the necessary training to develop appropriate teaching strategies in order to achieve different objectives depending on the school context but also will allow them to better understand their profession and the sense of it. The strategies used should be as flexible as possible, to suit specific circumstances and based, if necessary, on the experience of the teacher. Training, experience and creativity (alongside other elements such as collaboration, trust, etc.) are therefore, in the current educational context, the basic pillars on which to build this building.

"Mes amis, retenez ceci, il n'y a ni mauvaises herbes ni mauvais hommes. Il n'y a que de mauvais cultivateurs [My friends, stay with this: there is neither weeds nor bad men. There is only bad cultivators]" said Victor Hugo. Heed the words of the famous writer to focus on one final thought we only wish to target: the fact that it should be clear that the strategies the teacher uses in the classroom do not start from training in any way, but basically from their creativity. This is because education is not only a science or technology, but it is mainly (fortunately and sufficiently demonstrated by the best teachers) an art, as Comenius advocated. The teacher is also (and especially) a craftsman whose profession is possibly as close (if not more) to fine arts like cinema and literature as psychology or sociology. And there will always be teachers who create innovative and effective teaching strategies, as a result of passion and creativity, well above the various academic techniques that can be learnt.

It is true that the knowledge of these techniques is necessary, but in no way it can be sufficient for success and quality of education. I usually give my students the following example. Imagine that I am an art lover, specifically painting, and I marvel when I visit a museum or art gallery. Suppose I want to be like those painters. The first we have to learn are pictorial techniques. Learning oil painting, watercolor, how to use a spatula, different brushes, how to mix colors, the secrets of chromatics, etc. I could be in the best workshops in the world, even travel in time and visit Verocchio and Leonardo da Vinci as a teenager, learn the amazing technique of *sfumato*. I could be learning for ten, fifteen, twenty years, all my life, and I can assure that despite this learning and the mastery of these techniques, I won't ever be able to paint *The Sunflowers* by Van Goth, or *The Creation of Adam* by Miguel angel, or *The Meninas* by Velazquez, or *The Dog* by Goya. The question is that in order to paint these masterpieces it is not only necessary to know about painting techniques but, above all, to have been born for it, have a special gift that only a chosen few possess.

The same applies to education. There are teachers who have been born for it, people born with a gift, with that gift. Pure professionals whose teaching strategies come from the heart and not just their mind, wonderfully integrated in the classroom and not only capturing the attention of students but making them better people. Human beings who always bear the stamp of that male or female teacher and whom, throughout their lives, they will admire and, undoubtedly, will mark their existence. As Horace said "quod semel est imbuta recens, servabit odorem testa diu [the cask will long retain the flavor of that with which it was first filled]". Therefore, in this situation, the definitive question must be asked: today, does educational research respond to these needs? In other words, does educational research really help teachers?

## 8. BIBLIOGRAPHIC REFERENCES